\documentclass[twocolumn,pra,superscriptaddress,showpacs,aps,groupedaddress,floatfix]{revtex4-1}
\usepackage{graphicx}
\usepackage{dcolumn}
\usepackage{bm}
\usepackage{bbold}
\usepackage{stmaryrd}
\usepackage{latexsym}
\usepackage{amssymb}
\usepackage{amsfonts}
\usepackage{amsmath}
\usepackage{mathtools}
\usepackage{fancybox}
\usepackage{color}
\usepackage[breaklinks=true,colorlinks,citecolor=blue,linkcolor=blue,urlcolor=blue]{hyperref}
\usepackage[capitalize]{cleveref}
\usepackage{epstopdf}
\usepackage{braket}
\usepackage{csquotes}
\usepackage{ulem}
\usepackage[straightquotes]{newtxtt}

\begin{document}
\title{Dispersion forces between  weakly disordered van der Waals crystals}
\author{Jonas von Milczewski}
\email{jvmilczewski@mpq.mpg.de}
\affiliation{Max-Planck-Institute of Quantum Optics,
	Hans-Kopfermann-Stra{\ss}e 1, 85748 Garching, Germany}
\affiliation{Munich Center for Quantum Science and Technology,
	Schellingstra{\ss}e 4, 80799 Munich, Germany
}
\affiliation{Institute for Theoretical Physics, ETH Zurich, 8037 Zurich, Switzerland}
\author{John R. Tolsma}
\affiliation{Institute for Theoretical Physics, ETH Zurich, 8037 Zurich, Switzerland}
\date{\today}
\begin{abstract}
We describe a many-body theory for interlayer dispersion forces between weakly disordered atomically thin crystals and numerically investigate the role of disorder for different layer-separation distances and for different densities of induced electrons and holes. In contrast to the common wisdom that disorder tends to enhance the importance of Coulomb interactions in Fermi liquids, we find that short range disorder tends to {\it weaken} interlayer dispersion forces. This is in line with previous findings that suggest that transitioning from metallic to insulating propagation weakens interlayer dispersion forces. We demonstrate that disorder alters the scaling laws of dispersion forces and we comment on the role of the maximally crossed vertex-correction diagrams responsible for logarithmic divergences in the resistivity of two-dimensional metals.
\end{abstract}
%
%
\maketitle
\section{Introduction}
\label{sect:Intro}
Even when two objects are each electrically neutral, forces between the two objects which are mediated by the electromagnetic field can still be present. These {\it dispersion} forces were named by  London in his theoretical investigation of forces between molecules \cite{London_1930}. Although each molecule has zero total charge, quantum fluctuations in the charge density of each molecule lead to an effective dipole-dipole intermolecular force. This mechanism was later generalized by Lifshitz~\cite{Lifshitz1956,Dzyaloshinskii1961} to describe forces between solids, wherein he discovered a force which scales like $1/d^3$ when the distance $d$ between two thick slabs becomes large. Depending on the context, these forces also go under the name of van der Waals or Casimir forces, where the former (latter) often indicates that the force is mediated by the longitudinal (transverse) component of the electromagnetic gauge field~\cite{Jackson_EandM_book}. 

Dispersion forces are relatively weak and short ranged compared to electrostatic forces, and are difficult to observe in experiments on solids. Recently however, advances in x-ray spectroscopy have allowed for atomic-level precision measurements of interlayer strain in thin films and atomically thin crystals~\cite{Kozina_2014,ChengTung_2019}, and signatures consistent with interlayer dispersion forces among optically induced electrons and holes have been measured in transition-metal dichalcogenide multilayers~\cite{Mannebach_NanoLetters}. This adds a new, experimentally measurable quantity to the class of phenomena which are sensitive to correlations among quasiparticles in neighboring layers of atomically thin crystals like transition-metal dichalcogenides, graphene, twisted bilayer graphene, and phosphorene. Coulomb drag~\cite{Narozhny2016} is a notable example of the type of phenomena which are sensitive to interlayer correlations. In these experiments a current is driven in one layer and as a result of interlayer Coulomb interactions an induced voltage drop appears in a second (otherwise passive) nearby layer. Drag experiments have led to a deeper understanding of the nature of the elementary excitations and ground state wave functions of complex phases of matter, from two-dimensional Fermi liquids to more exotic phases like exciton condensates~\cite{Eisenstein2004,Kellogg2002} and Luttinger liquids~\cite{Laroche2014}. Just like Coulomb drag, the interlayer dispersion force between atomically thin crystals offers an interesting test bed for the various many-body theories describing the complex behavior of solids. 
\begin{figure}[t]
	\begin{center}
		\includegraphics[width=.80\linewidth]{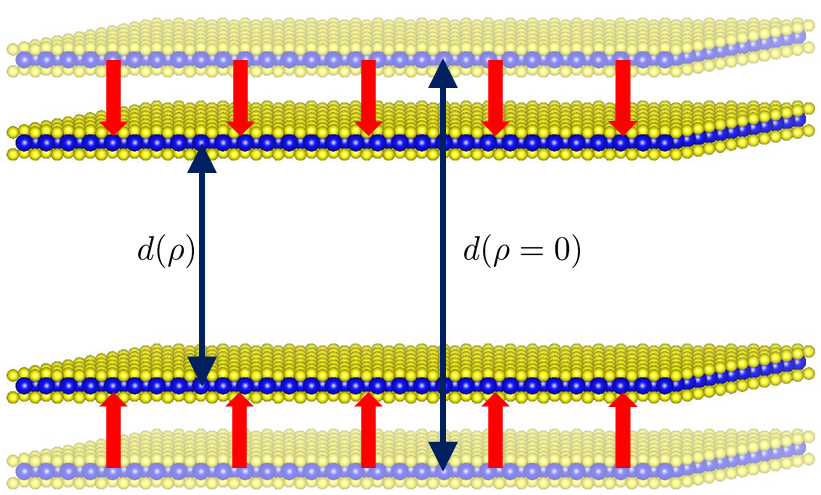} 
	\end{center}
	\caption{An illustration of the change in interlayer separation distance $d(\rho) - d(0)$, which results from the attractive forces between layers that are induced by  creating a finite density of electrons and holes in each layer, $\rho$.} \label{opener} \end{figure}

In this paper we construct many-body approximations to explore the impact of weak disorder on the interlayer dispersion forces which act between layers of a bilayer heterostructure after a finite density of electrons and holes are induced in each layer as illustrated in \cref{opener}. While {\it ab initio} methods for obtaining van der Waals contributions to the ground state energy exist~\cite{Andersson1996,Rydberg2003,Antony2006,Grimme2007,Tkatchenko2009,Tkatchenko2012}, our diagrammatic approach is sensitive to the exchange-correlation effects which density-functional theory usually deals with only on a mean-field level using variations of the local-density approximation; the approach discussed in this paper is complementary to these existing tools and allows for the treatment of systems with strongly correlated ground states or, as we investigate in detail below, random disorder. Quasiparticle-impurity interactions are known to be responsible for a number of fascinating properties of metals, from weak-localization corrections to the longitudinal conductivity~\cite{Gorkov1979} to anomalies in the tunneling conductivity~\cite{Altshuler1979a,Altshuler1979c}, and we will make use of some of these well-developed many-body approximations in determining the role of weak disorder on interlayer dispersion forces. 

Our paper is organized as follows. In Sec.~\ref{sect:bilayer} we describe a many-body theory for the interlayer dispersion force based on a linked-cluster expansion for the correlation energy of a bilayer in the absence of disorder as discussed previously \cite{Gramila1991,Sernelius1998}. In the limit of high quasiparticle density and large separation distance $d$, one recovers the well-known $d^{-5/2}$ scaling behavior \cite{Sernelius1998,Bostrom2000,Dobson2001} in agreement with predictions from Quantum Monte Carlo methods \cite{Drummond2007}. In Sec.~\ref{sect:diffuson} we describe a leading-order-in-$1/\varepsilon_{\rm F}\tau$ theory for interlayer forces. We demonstrate that disorder qualitatively alters the scaling laws and demonstrate that disorder tends to {\it reduce} the magnitude of interlayer forces. In Sec.~\ref{sect:cooperon} we discuss the impact on interlayer forces by a class of Feynman diagrams known to yield logarithmic divergences in the longitudinal resistivity of two-dimensional metals. Finally, in Sec.~\ref{sect:conclusion} we summarize our results and discuss interesting questions to be addressed in the future.

\section{Induced dispersion forces in bilayer systems}
\label{sect:bilayer}
We consider a system governed by the following Hamiltonian:
\begin{equation}
\mathcal{H}  = \mathcal{H}_0 + \mathcal{H}_{\text{e-e}} +\mathcal{H}_{\text{e-imp}} 
\end{equation}
which describes the kinetic energy of electrons and holes, the Coulomb interaction, and the interaction of electrons and holes with impurities, respectively. We assume, as is often the case experimentally, that the density of  induced electrons and holes (quasiparticles)  is such that the kinetic energy of electrons and holes can be described by an effective mass approximation
\begin{equation}
\mathcal{H} _0 = \sum_{{\bm k} \alpha I} \varepsilon_{\alpha}({\bm k}) \,\hat{a}^{\dagger}_{{\bm k} \alpha I}  \, \hat{a}^{\phantom{\dagger}}_{{\bm k}\alpha I} \ ,
\end{equation}
where $\varepsilon_{\alpha}({\bm k}) = \hbar^2 k^2/2m_{\alpha}$ and $\alpha$ is a composite index which labels the spin, valley, and band (e.g., valence vs conduction band) quantum numbers. In the following, we will consider the limit in which interlayer hopping is weak compared to the exchange-correlation energy per electron. Thus, the single-particle wave functions have a which-layer quantum number denoted by $I$. Interlayer hybridization of the conduction and valence bands is notoriously weak in van der Waals crystals (as the name suggests) and is often further weakened by rotational misalignment of neighboring layers.

The charged quasiparticles in the various layers of the system interact with each other via the Coulomb interaction
\begin{align}
\mathcal{H} _{\text{e-e}}  = \frac{1}{2 L^2}\sum_{\substack{{\bm q} \, I J \\ {\bm k}_1 {\bm k}_2  \\ \alpha \beta }}V_{I J}({\bm q}) \ \hat{a}^{\dagger}_{{\bm k}_1+{\bm q} \alpha I }  \hat{a}^{\dagger}_{{\bm k}_2-{\bm q} \beta J }  \hat{a}^{\phantom{\dagger}}_{{\bm k}_2 \beta J} \hat{a}^{\phantom{\dagger}}_{{\bm k}_1 \alpha I} \ , 
\end{align}
where 
\begin{equation}
V_{I J}({\bm q}) = \begin{cases} 
2 \pi e^2/(\kappa q), & I = J\\
2 \pi e^2 e^{-q d}/(\kappa q), & I \neq J.
\end{cases}\label{eq:twofields}
\end{equation}
The material-specific parameter $\kappa$ describes the dielectric contributions of the elementary excitations outside of our model [e.g., phonons and propagation of electric field outside of the two-dimensional (2D) material]. The strength of Coulomb interactions is traditionally \cite{Giuliani_and_Vignale} described by the value of a parameter $r_{s}$ which expresses the ratio of average interaction energy to average kinetic energy in a disorder-free two-dimensional electron gas (2DEG), $r_{s} \propto \langle\mathcal{H} _{\text{e-e}} \rangle / \langle\mathcal{H} _{0} \rangle $. The parameter depends on the total density of electrons (and holes) in each layer $n_I$ and is larger when the density is lower, $r_{s} = \left[a^*_B \sqrt{\pi n_I}\right]^{-1}$. Here, $a^*_B = \kappa a_B/m_{eff}$ is the effective Bohr radius. When the system contains particle populations described by different effective masses it is useful to define $a^*_B$ using the geometric mean of the masses $m_{eff} \rightarrow \sqrt{m_e m_h}$. Interactions of charged quasiparticles in different layers are ultimately responsible for the induced van der Waals forces we describe. In this paper we consider densities of induced quasiparticles which are large enough to form electron liquids and hole liquids rather than excitons, as was recently demonstrated at room temperature~\cite{Arp2019}.

The interaction between impurities of the crystal and electrons as well as holes is obtained by assuming that each impurity creates a deviation in the perfectly periodic scalar potential created by the underlying lattice. This scalar potential couples linearly to the density of electrons and holes, 
\begin{equation} \label{eq:Heimp}
\mathcal{H} _{\text{e-imp}} = \frac{1}{L^2}\sum_{{\bm Q}, I} u_I({\bm Q}) \rho_{I}({\bm Q}) \sum_{{\bm k} \alpha} \,\hat{a}^{\dagger}_{{\bm k}+{\bm Q} \alpha I}  \, \hat{a}^{\phantom{\dagger}}_{{\bm k}\alpha I} ,
\end{equation}
where $\rho_{I}({\bm Q})$ is the Fourier transform of the density of impurities in layer $I$, and $u_I({\bm Q})$ is the Fourier transform of the scalar potential of each impurity. We assume that electrons and holes only scatter off the impurity potential in the same layer, and we assume that the scalar potential is short ranged so that $u_I({\bm Q})$ is actually independent of wave vector. The quasiparticle-impurity scattering time $\tau_k$ can be defined using the Born approximation for the self-energy~\cite{Mahan} where $\Sigma({\bm k},\omega) = -i \hbar/2 \tau_{\bm k}$. In the presence of finite disorder, the scattering rate at the Fermi energy is used to define the small parameter of our perturbation theory $1/(\tau \varepsilon_{\rm F}) \ll 1$, where we here (and will continue to) drop the subscript on $\tau$.

Our method for evaluating the force between two atomically thin crystals consists of first calculating the ground state energy per layer as a function of interlayer separation distance $d$, and then calculating the force by taking the first derivative
\begin{equation}
\mathcal{F} = - \frac{1}{2}\frac{\partial E}{\partial d}\ .
\end{equation}
The ground state energy can be evaluated by taking the zero-temperature limit of the thermodynamic free energy  $\Omega$. The latter has a well-known perturbative formulation in the linked-cluster expansion~\cite{Mahan}

\begin{equation}\label{eq:FreeEnergy}
\begin{array}{l}
{\displaystyle \Omega-\Omega_0 = -\frac{1}{\beta}\sum_{\ell > 0} \frac{1}{\ell!}\left(\frac{-1}{\hbar}\right)^{\! \! \ell} \! \! \!  \int^{\hbar \beta}_0 \! \! \! d\tau_1 \ldots \int^{\hbar \beta}_0 \! \! \! d\tau_{\ell}}\vspace{0.2 cm}\\
{\displaystyle \quad \quad \quad \quad \times \, \, \,   {\rm tr} \left\{ \rho_0 {\rm T}_{\tau} \left[\hat{V}(\tau_1) \ldots \hat{V}(\tau_{\ell})\right] \right\} }_0 ~,
\end{array} 
\end{equation}
where $\rho_0$ is the noninteracting density matrix, ${\rm T}_{\tau}$ is the (imaginary) time-ordering operator, and $\hat{V}(\tau) = \mathcal{H} _{\text{e-imp}}(\tau) +  \mathcal{H} _{\text{e-e}}(\tau) $ is the sum of the two interactions in our model within the interaction picture of time evolution~\cite{Mahan}. By applying Wick's theorem, all contributions at order $\ell$ can be expressed in terms of integrals over noninteracting Green's functions, the Coulomb interaction $V$, and the electron-impurity interaction $u_I $. One can now make use of Feynman diagram techniques to efficiently calculate these contributions. We now have all the tools necessary to evaluate the interlayer force to any order in perturbation theory. 

\begin{figure}[t]
	\begin{center}
		\includegraphics[width=.90\linewidth]{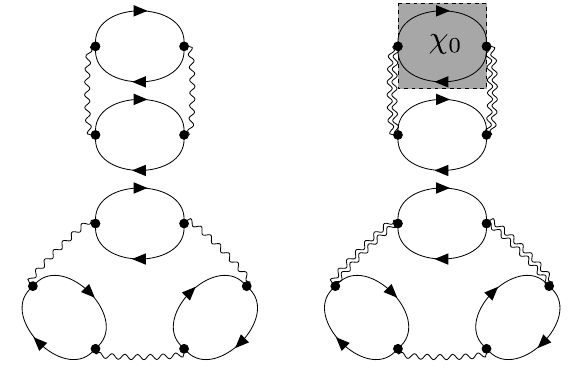}
	\end{center}
	\caption{Feynman diagrams for the correlation energy of a bilayer system whose quasiparticles interact via intralayer Coulomb interactions (single wavy lines) and interlayer Coulomb interactions (double wavy lines). Only the four lowest-order diagrams are shown here. Solid lines with arrows represent noninteracting Green's functions of quasiparticles. } \label{fig:RPAdiagrams} 
\end{figure}

Before we consider the effects of weak disorder on the interlayer forces, we reproduce the well-known $d^{-5/2}$ scaling of energy \cite{Tan1983,Sernelius1998,Bostrom2000,Dobson2001,Drummond2007}   by examining the force between two two-dimensional electron gases within the random-phase approximation (RPA) \cite{bohm1951collective,pines1952collective, bohm1953collective,pines1953collective} and taking the limit of large interlayer distance $d$. We thus ignore disorder and take $\hat{V}(\tau) = \mathcal{H} _{\text{e-e}}(\tau) $ within Eq.~(\ref{eq:FreeEnergy}). The RPA can be understood as an expansion of the ground state energy in powers of the small parameter $r_s$, and therefore gives a criterion for selecting which subset of Feynman diagrams at each order in $\ell$ within Eq.~(\ref{eq:FreeEnergy}) must be included in an approximation to a given order in $r_s$. The four lowest-order diagrams which contribute to the correlation energy are shown in Fig.~(\ref{fig:RPAdiagrams}). The full RPA approximation consists of summing all diagrams of this type, which at each order in $\ell$ contain $\ell$ bubble subdiagrams. The degeneracy of the diagrams in Fig.~(\ref{fig:RPAdiagrams}) is such that the infinite series of these types of diagrams can be resummed into a logarithm of a simple function of the single bubble diagram. After taking the derivative of the RPA approximation for the correlation energy  \cite{Sernelius1998}, we obtain the following integral expression for the force per layer between a bilayer system containing a finite density of electrons and holes in each layer
\begin{equation}
\mathcal{F}=-\frac{\hbar L^2}{4\pi^2} \! \! \int_0^{\infty} \! \! \! \! \, dq   \int_0^{\infty} \! \! \! \! \, d\omega \frac{q^2 V_{12}^2   \chi_0^2 }{	( 1-V_{11}\chi_0 )( 1-V_{22}\chi_0 )- V_{12}^2 
	\chi_0^2 }\  .  \label{eq:pressure}
\end{equation}
Here, $\chi_0$ is represented by the bubble subdiagrams found in the four diagrams in Fig.~(\ref{fig:RPAdiagrams}) and describes the noninteracting density-density response function of each layer. The zero-temperature limit of $\chi_0$ can be evaluated for parabolic-band effective mass models, and in the presence of both valence and conduction bands, $\chi_0=\sum_{\alpha}\chi_0^{\alpha}$, where $\chi_0^{\alpha}$ is the Lindhard function \cite{lindhard1954properties} of the $\alpha-$particle species. The integral over frequency in Eq.~(\ref{eq:pressure}) is over the imaginary frequency axis, and the arguments of $\chi_0^{\alpha}(q,i\omega)$ have been omitted for brevity. 

The application of Eq.~(\ref{eq:pressure}) assumes that thermal equilibrium has been reached among the electrons and holes, which is usually several orders of magnitude faster than the electron-hole recombination time, and does not limit experimental observations. For arbitrary electron/hole densities and interlayer separation distances, Eq.~(\ref{eq:pressure}) must be evaluated numerically. Furthermore, it should be mentioned that Eq.~(\ref{eq:pressure}) leads to a nonvanishing force even in the absence of holes. 

\begin{figure}[h]
	\begin{center}
		\includegraphics[width=.99\linewidth]{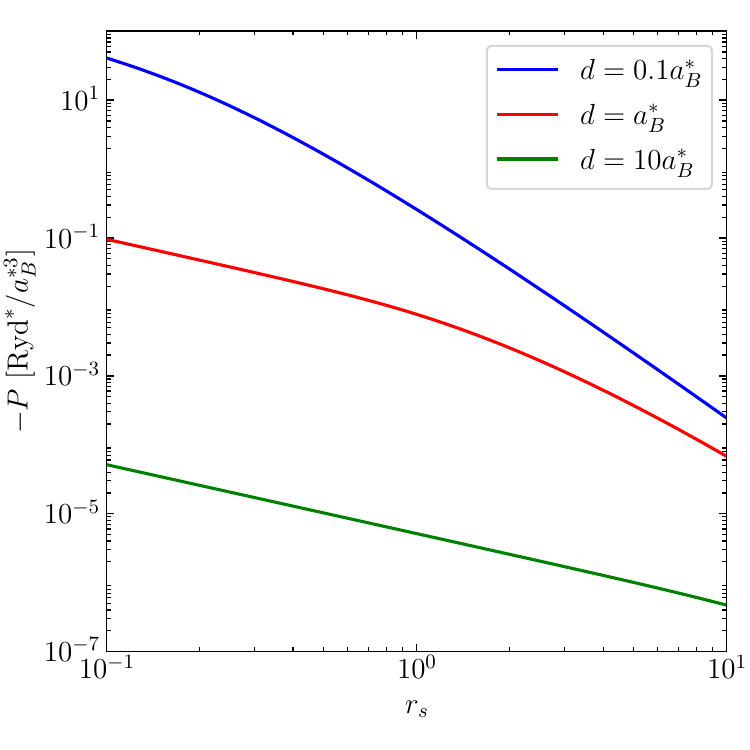} 
	\end{center}
	\caption{A plot of interlayer forces vs the density of induced quasiparticles in a disorder-free bilayer system. On the vertical axis is the force per area in units of effective Rydbergs per effective Bohr radius cubed. On the horizontal axis is the dimensionless parameter $r_s$ which is inversely proportional to the square root of the density of  induced quasiparticles. Explicit definitions for $r_s$  and $a_B^*$ can be found in the main text, while ${\rm Ryd}^*= e^2/ \kappa a_B^*$. } \label{fig:RPApressure} 
\end{figure}

In Fig.~(\ref{fig:RPApressure}) we present the results of numerical calculations for the pressure (i.e., force per area) between two layers of atomically thin crystals with induced densities of electrons and holes parametrized by $r_{s}$. We immediately notice that the force between layers is attractive and that the magnitude varies dramatically with interlayer separation distance  as $V_{12}$ is dependent on $d$. This is a particular feature of the type of dispersion force that derives from the instantaneous Coulomb interaction (typically called van der Waals forces) instead of forces originating from the transverse and retarded parts of the electromagnetic field (typically called Casimir forces). While Casimir forces act at larger distances than van der Waals forces, they are significantly weaker and they are independent of the amount of impurities in the materials, and therefore are not addressed in this paper.

To demonstrate how the RPA theory obtains the known $d^{-5/2}$ scaling for the energy \cite{Tan1983,Sernelius1998,Bostrom2000,Dobson2001,Drummond2007}, Eq.~(\ref{eq:pressure}) is now evaluated in the limit of large interlayer separation. Specifically, we will find the leading-order contribution to the interlayer force in the small parameter $1/(k_{\rm F} d)$, where $k_{\rm F}=\sqrt{k_{\rm F}^e k_{\rm F}^h}$ is the Fermi wave vector of the electron  and hole Fermi seas which are present in each layer after  excitation and thermalization. The presence of $e^{-2 q d}$ in the numerator of Eq.~(\ref{eq:pressure}) restricts the relevant range of $q$ in the integral to $q \lesssim 1/d$, which bears the physical interpretation that 2D
in-plane charge perturbation waves at wavelengths which are short compared to the interlayer distance appear averaged out on the adjacent plate and thus will not contribute
to forces. Long wavelengths, however, will not appear as averaged out and will therefore contribute to interlayer forces. In the limit $k_{\rm F} d\gg1$, the dominant contribution to
interlayer forces will then come from long in-plane wavelengths and this thus restricts
the relevant part of phase space to small values of $q$. In this region of phase space we are permitted to approximate $\chi_0^{\alpha}$ by its {\it dynamic} long-wavelength limit (i.e., $\omega>q$, $q \rightarrow 0$) which gives the leading-order contribution to the force. In the {\it dynamic} long-wavelength limit the noninteracting density-density response function of band $\alpha$ is given by
\begin{equation}
\chi_0^{\alpha} (q, i\omega)=  -\frac{\rho_{\alpha}}{m_{\alpha}} \frac{q^2}{\omega^2}\label{eq:BareChi} \ ,
\end{equation}
where $\rho_{\alpha}$ is the two-dimensional density of charged quasiparticles in band $\alpha$. It is then straightforward to evaluate Eq.~(\ref{eq:pressure}) analytically to obtain the leading order in $1/(k_{\rm F} d)$:
\begin{equation}
\mathcal{F}_{7/2}= - \frac{\hbar e \xi_1 L^2}{8\sqrt{2 \pi m} } \left( \frac{ \sqrt{\rho}}{d^{7/2}} \right) \ , \label{eq:analyticBilayer}  
\end{equation}
which corresponds to the $d^{-5/2}$ scaling for the energy. Here, $\xi_1 \approx 0.315$, $\rho$ is the total two-dimensional quasiparticle density in each layer, and we have taken $m_h=m_e=m$, and $\kappa=1$ for simplicity.  Interestingly, in the case of infinitely many parallel plates (superlattice), the scaling of force per layer  is identical to \cref{eq:analyticBilayer} up to redefinition of $\xi_1$ \cite{Mannebach_NanoLetters}. By randomly choosing two adjacent plates and identifying the gap between them as the gap between two semi-infinite thick slabs separated by a distance $d$, one can connect this result to Lifshitz' theory for thick, semi-infinite slabs. Introducing the three-dimensional density $\rho_{3D} = \rho/d$ in Eq.~(\ref{eq:analyticBilayer}) to compare with Lifshitz' theory, we immediately see that we have reproduced the power law for the interlayer force in terms of interlayer separation and quasiparticle density (i.e., $\mathcal{F} \propto \sqrt{\rho_{3D}} d^{-3}$).   

\begin{figure}[h]
	\begin{center} 
		\includegraphics[width=.92\linewidth]{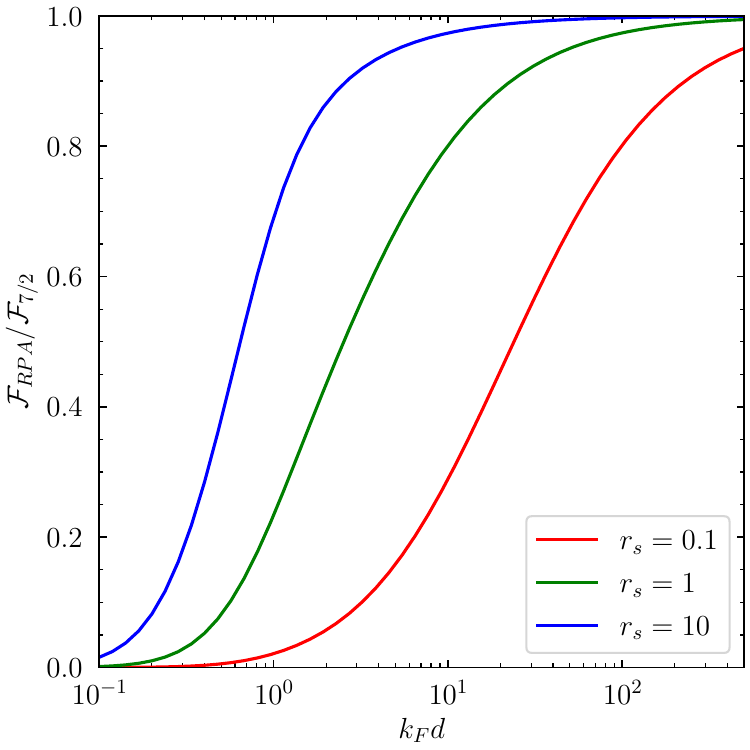} 
	\end{center}
	\caption{Ratio of the interlayer force in the random-phase approximation (RPA) $\mathcal{F}_{RPA}$, calculated numerically using Eq.~(\ref{eq:pressure}), to the interlayer force in the leading-order-in-$(k_{\rm F} d)^{-1}$ approximation $\mathcal{F}_{7/2}$, given by Eq.~(\ref{eq:analyticBilayer}). The more accurate RPA approximation predicts much smaller interlayer attraction unless $k_{\rm F} d \gg 1$, in which case both approximations yield the same result.} \label{fig:RPAoverLandau} 
\end{figure}

Despite the obvious utility of simple formulas like Eq.~(\ref{eq:analyticBilayer}), the derivation demonstrates that only the long-wavelength excitations (i.e., plasmons) are accounted for, while finite $q$ excitations (e.g., noncoherent particle-hole excitations) are neglected. Indeed, Eq.~(\ref{eq:analyticBilayer}) is only reasonable in the limit $1/(k_{\rm F} d) \ll 1$, and outside of this regime the interlayer forces are more accurately described by numerically evaluating Eq.~(\ref{eq:pressure}). This is demonstrated in \cref{fig:RPAoverLandau}, where we show the ratio of pressure in the RPA approximation of \cref{eq:pressure} to the asymptotic form of \cref{eq:analyticBilayer}. In the limit of $k_F d \gg 1 $ the predictions coincide, while for smaller values of $k_F d$ the asymptotic form gives much higher  interlayer attraction than the RPA form. In subsequent sections we will describe how these power law scalings are altered by the presence of impurities.

\section{Impact of disorder on VdW forces: The \texttt{'}Diffuson\texttt{'}}
\label{sect:diffuson}

In this section we lay out the basic elements of a many-body theory for the impact of weak disorder on the interlayer van der Waals (VDW) forces between atomically thin crystals. Specifically, we begin by introducing the small parameter  (i.e., $1/\varepsilon_{\rm F}\tau$) of the electron-impurity and hole-impurity interactions within the context of the first-order Born approximation (1BA) for the self-energy. We then identify the most relevant Feynman diagrams which contribute to interlayer dispersion forces within the regime of $r_s < 1/ \varepsilon_{\rm F}\tau$. These diagrams contain an infinite series of ladder diagrams, and we discuss the solution of the Bethe-Salpeter equation for the vertex correction of the density-response function in the limit of short ranged impurity potentials. In contrast to the effect of disorder on other phenomena which arise due to interlayer interactions (e.g., Coulomb drag \cite{Zheng1993}), we find that disorder tends to {\it weaken} the magnitude of van der Waals forces.

\begin{figure}[t]
	\begin{center}
		\includegraphics[width=.76\linewidth]{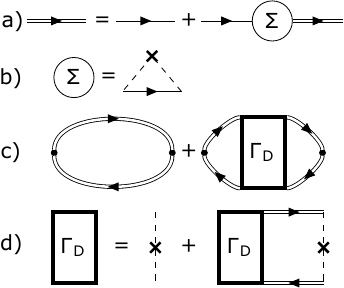} 
	\end{center}
	\caption{Feynman diagrams \cite{Ellis2017,Dohse2018} for the leading-order in $1/\varepsilon_{\rm F} \tau$ corrections to interlayer dispersion forces from impurity-quasiparticle interactions. Panel a) shows the diagrams for the Dyson equation for the self-energy. Single lines with arrows are noninteracting Green's functions and double lines with arrows are the noninteracting Green's functions dressed by scattering with impurities. Panel b) shows the proper self-energy in the first Born approximation. Each dashed line with a single cross represents the (disorder averaged) scattering off of the impurity potential. Panel c) shows the diagrams which contribute to the noninteracting density-response function in the leading-order in $1/\varepsilon_{\rm F} \tau$. Panel d) shows the diagrammatic representation of the Bethe-Salpeter equation for the {\it diffuson} contribution, {\it i.e.} the ladder-diagram vertex-correction $\Gamma_D (q,\omega)$} \label{fig:DiffusonDiagrams} 
\end{figure}

The electron-impurity and hole-impurity scattering rates can be defined by the 1BA for the self-energy. In this approximation the self-energy is purely imaginary, $\Sigma({\bm k},\omega) = -i \hbar/2 \tau_{\bm k}$. For simplicity, we will take the hole's and electron's impurity scattering rates to be equal, although this condition is easily relaxed if required. The 1BA is given by the Feynman diagrams depicted in panel a) and b) of Figs.~(\ref{fig:DiffusonDiagrams}). Explicitly, the 1BA for the $q$-independent scattering rate at the Fermi energy is 
\begin{equation}
\frac{1}{\tau} =  \frac{\nu_{\alpha} }{2 \hbar \pi}\rho^{imp} \vert u \vert^2 , \label{eq:1BAselfenergy}
\end{equation}
where $\nu_{\alpha}$ is the two-dimensional density of states at the Fermi surface of a single spin- and valley-resolved band, and $\rho^{imp} = \lim_{Q \rightarrow 0} \left[\rho_{I}({\bm Q}) \right]$. In obtaining Eq.~(\ref{eq:1BAselfenergy}) we have made two standard approximations for treating quenched disorder in solids \cite{Mahan}. First, we assume that the impurity potential is short ranged, such that the Fourier transform of the potential which appears in Eq.~(\ref{eq:Heimp}), $u_I({\bm Q})$, becomes independent of wave vector. Second, the impurity potential at any two different points is uncorrelated, such that the average over the probability distribution governing the impurity potential leads to $\left\langle \rho_I ({\bm Q}) \rho_I (-{\bm Q})\right\rangle_{imp} = N_{imp}$, where $N_{imp}$ is the number of impurities in layer $I$.

Next, we consider how to incorporate quasiparticle-quasiparticle interaction diagrams {\it and} quasiparticle-impurity interaction diagrams into an approximation for the dispersion force between atomically thin crystals. In the previous section we identified the leading-order-in-$r_s$ contribution to interlayer forces as the derivative of the RPA diagrams for the ground state energy. In order to work with a well-controlled perturbation theory we will restrict our selection of diagrams to the case when $r_s \ll 1/(\tau \varepsilon_{\rm F}) $. This allows us to obtain a well-controlled theory in both small parameters. The key is to not alter the order in $r_s$ of a diagram when adding any particular quasiparticle-impurity interaction line. We can accomplish this by adding to the RPA diagrams a nearly identical set of diagrams in which the noninteracting density-density response function bubble is {\it dressed} by quasiparticle-impurity interaction lines between the electron propagator and hole propagator which form each bubble.   As long as these {\it vertex-correction} quasiparticle-impurity lines do not cross each other, they can be summed to infinite order and together they give the leading order in $1/(\tau \varepsilon_{\rm F})$. The sum of all ladder Feynman diagrams for the density-density response function of each layer $I$ is represented in panels c) and d) of Fig.~(\ref{fig:DiffusonDiagrams}). The latter is the diagrammatic representation of the Bethe-Salpeter equation

\begin{equation}
\Gamma^{\rm D}_{{\bm k},{\bm k^{\prime}}}(q,\omega) = \Gamma_{{\bm k},{\bm k^{\prime}}}^0+ \sum_{{\bm k}^{\prime \prime}} \Gamma_{{\bm k},{\bm k^{\prime \prime}}}^0 \Pi_{\bm k^{\prime \prime}}(q,\omega) \Gamma^{\rm D}_{{\bm k^{\prime \prime}},{\bm k^{\prime}}}(q,\omega),
\end{equation}
where
\begin{equation}
\Pi_{\bm k^{\prime \prime}}(q,\omega) =  \frac{1}{\hbar^2 L^2}{\rm G}^{\rm R}({\bm k}^{\prime \prime}+{\bm q},\varepsilon_{\rm F}+\omega) {\rm G}^{\rm A}({\bm k}^{\prime \prime},\varepsilon_{\rm F} ) 
\end{equation}
and where ${\rm G}^{{\rm R}/{\rm A}}({\bm k},\omega) = \left[ \omega - \hbar^{-1} \xi_{{\bm k} \alpha} \pm i/2\tau\right]^{-1}$ and $\xi_{{\bm k} \alpha}  = \varepsilon_{{\bm k} \alpha} - \varepsilon_{{\rm F}}$. The Bethe-Salpeter  equation must usually be solved self-consistently for an arbitrary impurity potential. However, here it can be solved directly as a result of the bare-scattering amplitude being independent of momentum $\Gamma_{{\bm k},{\bm k^{\prime}}}^0 = \rho^{imp} \vert u_I \vert^2$. In the regime where disorder gives significant contributions to the density-density response of a system, $\omega < 1/\tau$ and $q < 1/v_{\rm F} \tau$, straightforward calculations  \cite{Vollhardt1980,Akkermans,Sadovskii2019} yield $\Gamma^{\rm D}(q,\omega) =\Gamma^0(q)/ \left[ -i \omega \tau + \tau {\rm D} q^2 \right] $ where the diffusion constant is defined in two dimensions as ${\rm D}=v_{\rm F}^2 \tau/2$. The diffusion pole present in $\Gamma^{\rm D}(q,\omega)$ at $\omega = -i {\rm D} q^2$ is also present in the disordered density-density response function of layer $I$ that is obtained by summing the diagrams in panel c) of Fig.~(\ref{fig:DiffusonDiagrams}) and yields
\begin{equation}
\chi_D (q, \omega)= -\nu_0 \frac{{\rm D} q^2}{-i \omega + {\rm D} q^2}, \label{diffusive}
\end{equation}
where $\nu_0$ is the total density of states at the Fermi energy in layer $I$. 

\begin{figure}[t]
	\begin{center}
		\includegraphics[width=.95\linewidth]{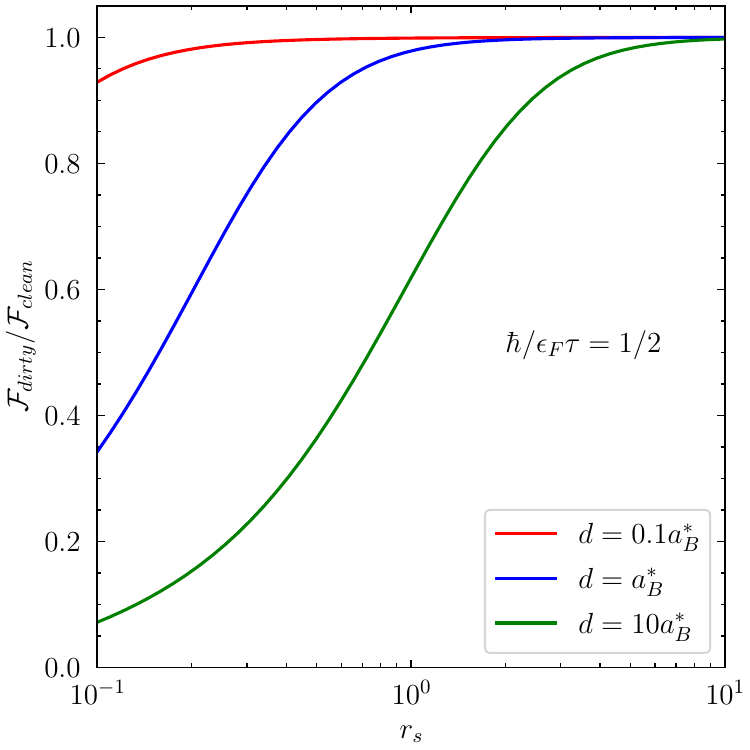} 
	\end{center}
	\caption{Ratio of the interlayer force in the presence of disorder $\mathcal{F}_{dirty}$ to the interlayer force with no disorder $\mathcal{F}_{clean}$, plotted against the interaction parameter $r_s$ which is inversely proportional to the square root of the induced quasiparticle density in each layer of a bilayer. The three curves are for three different values of the interlayer separation distance $d$ in units of the effective Bohr radius $a_B^*$. The degree of disorder is given by $\hbar/\epsilon_F \tau= 1/2$. The values of both $\mathcal{F}_{dirty}$ and $\mathcal{F}_{clean}$ are calculated numerically using Eq.~(\ref{eq:pressure}). In $\mathcal{F}_{dirty}$, the density response is given by the disordered limit $\chi_D(q,\omega)$  for $\omega < 1/\tau$ and $q < 1/v_{\rm F} \tau$.} \label{fig:Diffusion}  
\end{figure}

We can now evaluate the effect of weak disorder on the dispersion force between two atomically thin crystals by numerically evaluating Eq.~(\ref{eq:pressure}) after replacing $\chi_0(q,i\omega)$ by $\chi_D(q,i\omega)$ in the region of phase space where $\omega < 1/\tau$ and $q < 1/v_{\rm F} \tau$. In Fig.~(\ref{fig:Diffusion}) we plot the ratio of the interlayer force in the presence of disorder $\mathcal{F}_{dirty}$ to the force in the absence of disorder $\mathcal{F}_{clean}$. We find that the interlayer attraction is reduced in magnitude by the presence of quasiparticle-impurity interactions, which we will analyze in more detail below. We also find that $\mathcal{F}_{dirty}/\mathcal{F}_{clean}$ is reduced as $d$ increases. This occurs due to the presence of $e^{-2 q d}$ in Eq.~(\ref{eq:pressure}) which originates from the form of the 2D in-plane Fourier transform of the interlayer Coulomb interaction. This factor restricts the density fluctuations which contribute to interlayer forces to wave vectors $q \lesssim 1/2d$, and as $d$ is increased, more of this region of phase space lies in the region governed by the disordered density-density response, $q < 1/v_{\rm F}\tau$. We will now show that this phase space effect is also responsible for a change in the power-laws for the dispersion forces at large interlayer separation distances. In other words, in the presence of disorder, the asymptotic limit for forces between 2D planes presented in Eq.~(\ref{eq:analyticBilayer}), $\mathcal{F} \propto d^{-7/2}$, is altered. 

The numerical results presented in Fig.~(\ref{fig:Diffusion}) show that disorder {\it decreases} the magnitude of interlayer forces. This is in contrast to the effect of disorder on other phenomena, like Coulomb drag \cite{Zheng1993}, which also originates from interlayer quasiparticle-quasiparticle interactions. In the case of Coulomb drag, this conventional cartoon picture of the effect of disorder is that the change in the density-density response function from the noninteracting limit $\chi_0(q,i\omega)$ to the disordered limit $\chi_D(q,i\omega)$ represents a change from ballistic to diffusive motion of the quasiparticles. Indeed, the disordered density-density response function can be derived from semiclassical arguments using the diffusion equation \cite{Giuliani_and_Vignale}, which is equivalent to the relaxation time approximation (RTA)  \cite{Mermin1970} in the region $\omega < 1/\tau,\ q < 1/v_{\rm F} \tau$ in the {\it dynamic} limit. Since quasiparticles in neighboring layers which experience diffusive motion tend to spend longer periods of time near each other, they interact more strongly and this increases the Coulomb drag (i.e., disorder tends to enhance the transresistivity).  However, since the interlayer forces are decreased in magnitude by the presence of disorder, we find that the cartoon picture of the effect of disorder cannot be imported to understand our case of interest. The reason why disorder decreases interlayer forces while increasing the interlayer Coulomb drag is most simply identified by again examining the large-$d$ limit of the two quantities. Specifically, while both Coulomb drag and the interlayer force depend on the density-density response function, the leading-order-in-$1/(k_{\rm F}d)$ contribution to Coulomb drag comes from the {\it static} limit ($\omega<q$, $q \rightarrow 0$) of $\chi(q,i\omega)$ while the analogous contribution to the interlayer force comes from the {\it dynamic} limit ($\omega>q$, $q \rightarrow 0$) of $\chi(q,i\omega)$. 

In the large-$d$ limit our numerical results for the correlation energy per layer can be compared to previous investigations of disordered correlation energies within single-layer systems \cite{Asgari2002} where it was found that the introduction of disorder increases exchange energies in magnitude but decreases correlation energies in magnitude. By following similar steps as we took to derive the disorder-free expression presented in Eq.~(\ref{eq:analyticBilayer}), we find the following leading-order expression:
\begin{equation}
\mathcal{F}_{dirty}= -\frac{\hbar e^2 \xi_2 L^2 \tau }{4 \pi m} \left( \frac{\rho}{d^4} \right),
\end{equation}	
where $\xi_2 \approx  0.768$ and $\rho$ is the total two-dimensional density of quasiparticles in each layer and we have again taken $m_e=m_h=m,\ \kappa=1$ for simplicity. Notice that the interlayer force now decays more quickly with distance than in the absence of disorder. This qualitative change is a direct result of the transition of electron and hole propagation from ballistic to diffusive.

While it might be surprising that the effect of disorder on the interlayer forces is opposite to its effect on interlayer Coulomb drag, this behavior actually fits nicely into a trend observed in other systems \cite{Dobson2006,Dobson2012}:  the less metallic a system is, the faster its energy (and therefore its pressure) decreases with interlayer separation. Concretely, for a metallic sample, the energy scales as $d^{-5/2}$ \cite{Tan1983,Sernelius1998,Bostrom2000,Dobson2001,Drummond2007} while for a combination of a graphene and a metallic plate it scales as $\log(d) d^{-3}$ \cite{Dobson2006} and for two graphene plates is scales as $d^{-3}$ \cite{Dobson2006}. Finally, for two insulator system it scales as $d^{-4}$ \cite{Rydberg2003}. The change of the scaling of the distance-dependent part of the correlation energy from $d^{-5/2}$ to $d^{-3}$ upon changing from ballistic to diffusive propagation thus confirms this picture.

\section{Quantum interference effects  on VdW forces: The \texttt{'}Cooperon\texttt{'}}
\label{sect:cooperon}
In the previous section we developed a theory for interlayer dispersion forces between the layers of a bilayer system of atomically thin crystals which have uncorrelated and short ranged disorder. We summed an infinite set of Feynman diagrams by solving the Bethe-Salpeter equation and thus obtained the {\it diffuson} vertex correction of the density-density response function to leading-order in $1/(\epsilon_{\rm F}\tau)$. 
In this section we will sum the class of diagrams which corresponds to the subleading-order terms for the interlayer dispersion force in powers of $1/\varepsilon_{\rm F}\tau$. These diagrams are familiar from the theory of weak-localization and together they constitute the {\it cooperon} vertex-correction. Despite being of lower order in the small parameter governing the impurity-quasiparticle interaction, they are known to be responsible for a logarithmic divergence in the longitudinal resistivity of two-dimensional conductors \cite{Gorkov1979}, which motivates us to consider them here as well.
\begin{figure}[t]
	\begin{center}
		\includegraphics[width=.83\linewidth]{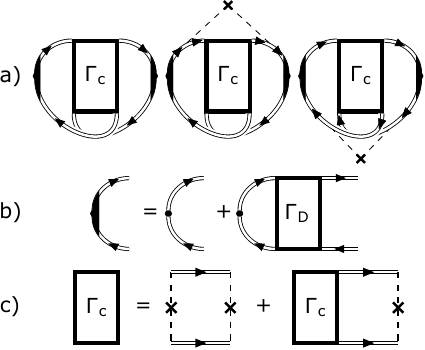} 
	\end{center}
	\caption{Feynman diagrams representing the $cooperon$ contributions to the noninteracting density-response function from scattering of electron and holes off of the impurity potential. Panel a) shows the three diagrams that contribute at subleading order in $1/\varepsilon_{\rm F}\tau$. Panel b) shows the diagrams describing the {\it diffuson} dressing of the density-fluctuation operator. Panel c) shows the Bethe-Salpeter representation of the maximally crossed diagrams that represent the vertex-correction $\Gamma_C(q,\omega)$.} \label{fig:CooperonDiagrams} 
\end{figure}

The cooperon contributions to the density-density response function are obtained by summing the 'maximally crossed' vertex-correction; this infinite set of diagrams is illustrated in panel c) of Fig.~(\ref{fig:CooperonDiagrams}). These diagrams represent the quantum interference of a wave packet of charge density which interferes with itself while traversing along the time-reversed path. This requires the system to have a time-reversal symmetry present in order for phase coherence to be maintained in-between collisions of the wave packet with different  impurities. As previously mentioned, these diagrams give a logarithmic divergence in the resistivity (which is proportional to the current-density response function), and indeed a similar phenomenon happens in our case of interest. Specifically, the subleading-order contribution to the density-density response function yields a logarithmic divergence in the diffusion constant. When both the diffuson and cooperon contributions to the density-density response function are included \cite{RaimondiBook}, the functional form of $\chi_D(q,\omega)$ remains the same as presented in the last section except that $\rm D$ gets an additional contribution which depends on frequency
\begin{equation}\label{eq:NewD}
\delta {\rm D}(\omega) = -\frac{1}{4 \pi^2 \hbar \nu_0} \log\left[ \frac{1+2\tau\omega}{\left( \tau/\tau_0 \right)^2 + 2 \tau \omega}\right] \ , 
\end{equation}
where $\nu_0$ is the total two-dimensional density of states of all quasiparticles in layer $I$. Just as in the case of the cooperon contribution to the longitudinal resistivity, the logarithmic divergence we obtain is cutoff at long distances, or small momenta, by the inelastic scattering time of the quasiparticles, $\tau_0$. This time-scale is determined, for example, by the quasiparticle-quasiparticle scattering rate, and is responsible for destroying the phase coherence of the propagating (and time-reversed propagating) wave packet on very long time scales  $\tau_0 > \tau$. This form of the disordered response function is only a reasonable approximation in the range where $\omega<1/\tau$ and $q < 1/v_{\rm F}\tau$. 

\begin{figure}[t]
	\begin{center}
		\includegraphics[width=.91\linewidth]{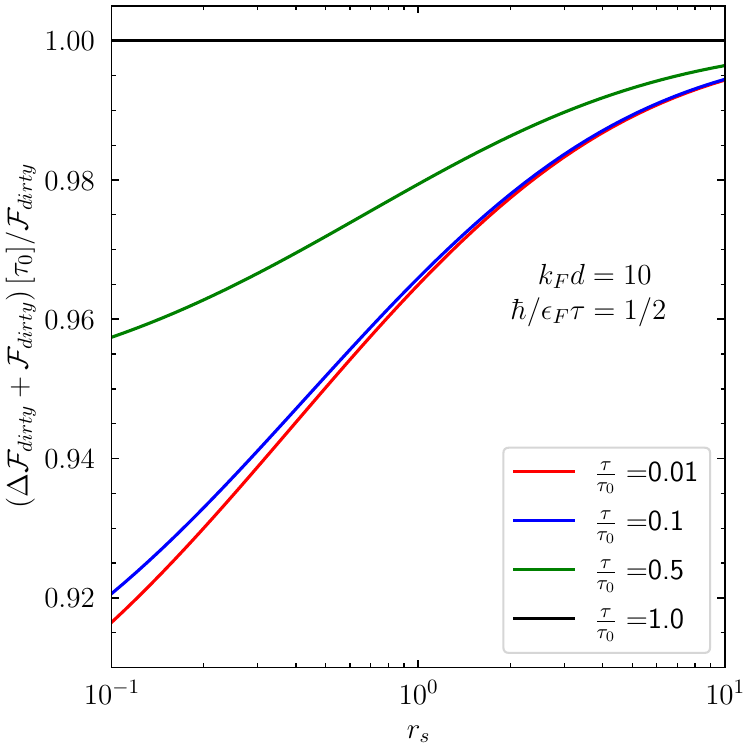} 
	\end{center}
	\caption{The fractional change in the interlayer dispersion force when the maximally crossed (i.e., weak-localization) diagrams are included. Notably, the logarithmic divergence which appears in the longitudinal resistivity of two-dimensional conductors is not present here. Instead, the {\it cooperon} diagrams have a similar, but weaker, effect as the {\it diffuson} diagrams, where both tend to reduce the magnitude of interlayer attractive forces. The fractional changes are shown at $\hbar/\epsilon_F \tau= 1/2$, $k_Fd=10$ as a function of $r_s$ for different values of $\tau/\tau_0$. } \label{fig:Cooperon} 
\end{figure}

We numerically evaluate the interlayer dispersion forces using the disordered density-density response function including the renormalized diffusion constant  ${\rm D} \rightarrow {\rm D} + \delta{\rm D}(\omega)$. The results are shown in Fig.~(\ref{fig:Cooperon}). They demonstrate that the maximally crossed diagrams tend to further reduce the magnitude of interlayer forces. More surprisingly, perhaps, there is no logarithmic divergence in the interlayer force, in contrast to what happens when using the analogous approximation for the longitudinal conductivity. This is surprising in light of the well-known relationship $\sigma_{\rm dc} = \lim_{q \rightarrow 0} (q^2/\omega^2) \chi_D(q,\omega)$, which follows from the presence of global gauge symmetry. However, while the conductivity is the response of the system to an external electric field whose frequency we can always fix to zero, in contrast, the interlayer dispersion force is an {\it integral over all frequencies} of density fluctuations in both layers (it is the Coulomb interaction between these density fluctuations which yields the dispersion force). And when the logarithmic divergence in $\chi_D(q,\omega)$ is integrated over frequency, it results simply in a finite reduction (on the order of $\lesssim 10$ percent)  of the interlayer force's magnitude.  
\section{Summary And Discussion}
\label{sect:conclusion}
We developed a many-body theory for the dispersion forces between atomically thin crystals with weak disorder. Such systems can be realized within van der Waals crystals \cite{Geim_Nature} (e.g., graphene, transition-metal dichalcogenides, etc.) which form multilayer systems with very weak interlayer hybridization, a property which has allowed for optically induced interlayer strain, originating from dispersion forces, to be observed recently \cite{Mannebach_NanoLetters}. In these systems {\it dispersion} forces arise due to Coulomb interactions between fluctuations in the charge density of neighboring layers. The linked-cluster expansion method was used to approximate the correlation energy of a bilayer system and the force between the layers of the bilayer system was obtained by taking a derivative of the correlation energy with respect to interlayer separation distance. 

In the high-density limit, the random-phase approximation bubble diagrams give the leading-order contribution to the disorder-free interlayer dispersion force. To account for disorder, we have summed an infinite series of ladder diagrams by solving the Bethe-Salpeter equation. These ladder diagrams form the {\it diffuson} contribution to the vertex correction of the density-density response function (i.e., the bubble), and yield the leading-order-in-$1/(\varepsilon_{\rm F} \tau)$ theory. Numerical evaluation of the interlayer dispersion force shows that  interlayer forces are {\it weakened} by disorder. On one hand, this is in contrast to the more conventional case \cite{Lee_ReviewModernPhysics_1985} in which Coulomb interactions become more important when electron motion becomes diffusive rather than ballistic.  On the other hand, this behavior is in accordance with previously observed changes in scaling laws as one transitions from metallic to insulating electron propagation. 

We explain this behavior by considering the analytic structure of the density-response function in the small frequency and wave-vector limit. We find that the diffusive motion of electrons and holes leads to a qualitative change in the scaling laws for the interlayer dispersion force as a function of quasiparticle density and interlayer separation distance. Subsequently, the impact of the higher-order vertex-correction diagrams was investigated. Specifically, maximally crossed diagrams which are known to produce logarithmic divergences in the longitudinal resistivity of two-dimensional metals (i.e., weak localization diagrams) are found to be much less important for interlayer dispersion forces. 

All the calculations shown in this paper were carried out within a bilayer system consisting of two parallel plates. It should be mentioned, however, that the effects of the theories developed in this paper were all at the level of \enquote{same-layer}-density-density response functions. As the theory of the bilayer system can easily be generalized to the theory of a superlattice system \cite{Mannebach_NanoLetters}, the results of this paper can easily be transferred to the superlattice system with similar effects (e.g., same power laws and qualitative effects). 

Optical control of electron and hole populations yields a convenient control knob for manipulating the interlayer separation distance of van der Waals crystals.  In future calculations one may investigate the possibility of inducing interlayer dispersion forces by doping heterostructures with electrostatic gates. While these systems include interlayer electrostatic forces which compete with dispersion forces, the latter are not reliant on equal populations of electrons and holes and can hopefully still be observed. Through electrostatic gating the role of the excitonic spectrum in the formation of strains could be differentiated from the  induced strains presented in our work. In order to complement this investigation of the role of excitons, it would furthermore be interesting to study the qualitative changes in interlayer dispersion forces which are present in multilayer systems with more exotic ground state wave functions, such as are present in bilayer exciton condensates. 
\begin{acknowledgments}
J. v. M. is supported by a fellowship of the International
Max Planck Research School for Quantum Science and Technology (IMPRS-QST). J. R. T. acknowledges financial support from the Swiss National Science Foundation.
\end{acknowledgments}

\end{document}